\documentclass[prb,twocolumn,showpacs,preprintnumbers,amsmath,amssymb]{revtex4}
\usepackage{graphics}

\begin{document}

\title{Norm-conserving Hartree-Fock pseudopotentials and their
 asymptotic behaviour}

\author{J. R. Trail}
\email{jrt32@cam.ac.uk}
\affiliation{TCM Group, Cavendish Laboratory, University of Cambridge,
 Madingley Road, Cambridge, CB3 0HE, UK}
\author{R. J. Needs}
\affiliation{TCM Group, Cavendish Laboratory, University of Cambridge,
 Madingley Road, Cambridge, CB3 0HE, UK}

\date{October, 2004}

\begin{abstract}
We investigate the properties of norm-conserving pseudopotentials
(effective core potentials) generated by inversion of the Hartree-Fock 
equations.  In particular we investigate the asymptotic behaviour
as $\mathbf{r} \rightarrow \infty$ and find that such pseudopotentials
are non-local over all space, apart from a few special special cases
such H and He. Such extreme non-locality 
leads to a lack of transferability and, within periodic
boundary conditions, an undefined total energy.  The extreme
non-locality must therefore be removed, and we argue that the best way
to accomplish this is a minor relaxation of the norm-conservation
condition.  This is implemented, and pseudopotentials for the atoms
H$-$Ar are constructed and tested.
\end{abstract}



\maketitle

The pseudopotential approximation is a vital part of the practical
application of \emph{ab initio} methods to problems in quantum
chemistry and condensed matter physics. In the pseudopotential 
approach the tightly bound core electrons are removed and their 
influence on the rest of the system is represented by an 
effective potential.  This reduces the number of electrons that 
must be considered and yields a smoother potential.  These 
properties significantly reduce the computational effort required 
for complex systems.

Over the past few years ultra-soft\cite{vanderbilt90,kresse94}
pseudopotentials and separable norm-conserving Kleinman-Bylander
\cite{kleinman82} pseudopotentials have come to dominate within
plane-wave Density Functional Theory (DFT) technology.  Within the
quantum chemistry community it is more usual to use non-separated
norm-conserving pseudopotentials, which are the kind we deal with
here.  Such pseudopotentials can, however, be converted to the
separable Kleinman-Bylander form.

Our main interest is in diffusion quantum Monte Carlo (DMC)
calculations~\cite{foulkes01}. Norm-conserving pseudopotentials are
routinely used within DMC calculations~\cite{non_local_qmc,mitas91},
and there is evidence to show that Hartree-Fock (HF) theory provides 
better pseudopotentials for this purpose than DFT~\cite{greeff98}. 
HF pseudopotentials have been used in
various correlated valence quantum chemistry calculations, and we
suggest they would be suitable for use in perturbation theories such
as the $GW$ scheme, which has been applied to a number of condensed
matter systems\cite{gw_examples}.

It turns out that most of the HF pseudopotentials available in the
quantum chemistry literature diverge at the origin, normally like
$1/r^2$ or $1/r$, which makes them unsuitable for use in DMC
calculations\cite{bachelet82,hay-wadt,christiansen79,dolg87}.
Recently Ovcharenko \textit{et al.}\cite{ovcharenko01} developed HF
pseudopotentials for Be-Ne and Al-Ar which are finite at the origin,
and we will compare the pseudopotentials generated here with these. We
would like our pseudopotentials to be as smooth as possible, since
lack of smoothness can reduce the transferability of a
pseudopotential. In addition we wish to generate HF orbitals 
using a variety of basis sets, including plane waves, for which 
a smooth pseudopotential reduces the computational cost.

It is advantageous to make the region over which the pseudopotential
is non-local as small as possible as the evaluation of the non-local
energy within DMC is expensive. Having a small region of non-locality
also tends to promote transferability, but tends to make the 
pseudopotential less smooth.

An important issue arises in generating HF pseudopotentials by
inversion of the HF equations.  It has long been understood that if
the norm is not conserved the Hartree potential due to the pseudo-ion
does not decay as $-Z_{val}/r$ at large $r$ (where $Z_{val}$ is the
ionic charge)\cite{krauss84}.  It has also been appreciated that the
exchange interaction gives a long ranged tail to the pseudopotential
which should be removed \cite{krauss84,christiansen79}. We find that
even if norm-conservation is enforced the exchange interaction results
in a ``non-Coulombic tail'' that decays more slowly than the Coulomb
part of the pseudopotential, persists far from the atom, and is
non-local.  Non-Coulombic tails in HF pseudopotentials have been reported
before in the literature, but the form that we derive here is
different from that described by Kahn \emph{et al.}\cite{kahn76} and
Hay \emph{et al.}\cite{hay78}, who suggest that the deviation from
$-Z_{val}/r$ decays faster than $1/r$.

These non-Coulombic tails increase the cost of calculations, lead to
unphysical results and, in the case of periodic boundary conditions,
leave the total energy undefined.  Consequently the tail must be
removed for practical applications. 
Although past workers have constructed HF pseudopotentials with this 
long range non-local tail removed, we are not aware of a detailed
discussion in the literature of the form and magnitude of this tail.
Understanding this long ranged effect is important when considering
how to remove it.

In what follows we investigate the general properties of
norm-conserving\cite{shirley89} HF pseudopotentials, with particular
emphasis on their non-locality and asymptotic behaviour.
We present a method to localise the potential such that it is non-local 
only within a relatively small region surrounding the nucleus, and 
enforce the desired asymptotic behaviour.

In section \ref{sec:def_pp} we discuss pseudopotentials obtained from
inversion of the HF equations, in section \ref{sec:nloc} we derive the
asymptotic behaviour of these pseudopotentials, and report detailed
results for Ne.  In section \ref{sec:results} we present and analyse 
results for the atoms H-Ar.
We draw our conclusions in section \ref{sec:conc}.

Atomic units are used throughout, unless otherwise indicated.

\section{Pseudopotentials from inversion of the HF equations}
\label{sec:def_pp}

The pseudopotential generation procedure we apply is similar to that
used by others, such as Hay and Wadt\cite{hay-wadt}, or Troullier
and Martins\cite{troullier91}. In this section we provide a brief
summary of the procedure for generating norm-conserving
pseudopotentials, primarily to define the notation and context.

For an isolated atom a HF orbital\cite{fischer97}, $\psi_{ilm}$, may
be labelled by the quantum numbers $i$, $l_i$ and $m$.  Separating the
angular and radial coordinates leads to
\begin{equation}
\psi_{il_im}(\mathbf{r})=\frac{\phi_{i}(r)}{r} Y_{l_im}(\Omega).
\label{eq:1.1}
\end{equation}
The radial part of the orbital, $\phi_{i}(r)$, satisfies
\begin{equation}
 \left[
- \frac{1}{2} \frac{d^2}{dr^2}
+ \frac{l_i(l_i+1)}{2r^2}
+ V^{eff}_i
 \right] \phi_{i} 
= \epsilon_{i} \phi_{i},
\label{eq:1.2}
\end{equation}
where the effective potential for orbital $i$ is
\begin{equation}
V^{eff}_i=
- \frac{Z}{r} 
+ V_h[ \rho ] 
+ \frac{ \hat{V}_x[ \{ \phi \},l_i ] \phi_i}{ \phi_i }.
\label{eq:1.3}
\end{equation}
The first term arises from the nuclear charge and the second is the
Hartree potential due to the total electron density.  The third term
is the effective exchange potential, which is different for each
orbital.  For convenience in what follows we have included the
(cancelling) self-interaction in the second and third terms.

To construct a pseudopotential we first partition the atomic orbitals
into core orbitals, whose influence will be represented by the
pseudopotential, and valence orbitals, which will be represented by
the pseudo-orbitals. The eigenvalues of the pseudo-orbitals 
are constrained to equal the equivalent all-electron (AE) 
eigenvalues, and the pseudo-orbitals themselves 
are constrained to equal the equivalent AE orbitals outside of a 
``core radius'', $r_{ci}$.  Inside $r_{ci}$ the orbitals are 
given by an as-yet unspecified function $f_i$, so that
\begin{equation}
\tilde{\phi}_i(r) =
  \left\{
  \begin{array}{ll}
    f_i(r)          & r <    r_{ci} \\
    \phi_i(r)       & r \geq r_{ci}\;.
  \end{array}
  \right.
\label{eq:1.4}
\end{equation}
$f_i(r)$ is chosen such that $\tilde{\phi}_i(r)$ is nodeless, 
smooth up to a certain order of differentiation at $r=r_{ci}$, 
and that its norm is equal to that of $\phi_i(r)$.

The pseudo-orbital for state $i$ satisfies
\begin{equation}
 \left[
- \frac{1}{2} \frac{d^2}{dr^2}
+ \frac{l_i(l_i+1)}{2r^2}
+ \tilde{V}^{eff}_i
 \right] \tilde{\phi}_{i} 
= \epsilon_{i} \tilde{\phi}_{i}\;,
\label{eq:1.5}
\end{equation}
and hence the effective potential, $\tilde{V}^{eff}_i$, can be
obtained by inverting Eq.~(\ref{eq:1.5}).  The pseudopotential for
state $i$, $\tilde{V}_{i}(r)$, is then defined by the pseudo-atom
equivalent of Eq.~(\ref{eq:1.3}):
\begin{equation}
 \tilde{V}^{eff}_i =
  \tilde{V}_{i}(r)
+ V_h[ \tilde{\rho} ] 
+ \frac{ \hat{V}_x[ \{ \tilde{\phi} \}, l_i]
     \tilde{\phi}_i }{ \tilde{\phi}_i }\;.
\label{eq:1.6}
\end{equation}

To use the pseudopotential in a calculation for a molecule or solid it
is expressed in terms of projection operators, and separated into
local and non-local parts,
\begin{eqnarray}
\hat{V}_{pseudo}&=&\tilde{V}_{local}(r) \nonumber \\*
   & & + \sum_{l}^{l_{max}} \sum_{m=-l }^{l} 
  | Y_{lm} \rangle ( \tilde{V}_{l}(r) - \tilde{V}_{local}(r)) 
   \langle Y_{lm} |. \nonumber \\*
\label{eq:1.7}
\end{eqnarray}
The orbitals with $l > l_{max}$ feel the local potential,
$\tilde{V}_{local}$.  Here the index $l$ is interchangeable with the
pseudo-atom orbital index, $i$, since the valence states chosen to
construct the pseudopotential must have unique $l$ quantum numbers.

\section{Extreme non-locality of HF pseudopotentials}
\label{sec:nloc}

In this section we address the question of the locality of the
pseudopotentials defined above.  We define a pseudopotential to be
local within a radius $r_{loc}$ if, for all $i$ and $j$,
\begin{equation}
 |\tilde{V}_i(r) - \tilde{V}_{j}(r)| \leq \delta {\textnormal{ for
  }} r > r_{loc},
\label{eq:2.1}
\end{equation}
for some $r_{loc}$ and small $\delta$.

By definition $\tilde{\phi}_i=\phi_i$ for $r>r_{ci}$ and hence
$\tilde{V}^{eff}_i = V^{eff}_i$ for $r>r_{ci}$.  This equality,
together with Eqs.~(\ref{eq:1.3}) and (\ref{eq:1.6}) gives
\begin{eqnarray}
 \tilde{V}_{i}(r) &=& - \frac{Z}{r} + V_h[ \rho - \tilde{\rho} ] 
     \nonumber \\*
 & &+ \frac{ \hat{V}_x[ \{ \phi \}, l_i ] \phi_{i} }{ \phi_{i} } 
- \frac{ \hat{V}_x[ \{ \tilde{\phi} \}, l_i ] 
\tilde{\phi}_{i} }{ \phi_{i} }, \;\;\;\; r>r_{ci}. \nonumber \\*
\label{eq:2.2}
\end{eqnarray}
The first two terms are independent of the orbital and, since 
$\rho$ and $\tilde{\rho}$ differ by the exponentially decaying core
contribution, the sum of these terms approaches $-Z_{val}/r$ as
$r\rightarrow\infty$.  However, since the HF exchange potential is
non-local there is no reason to expect the final two terms to cancel,
or to approach zero faster than $1/r$ as $r\rightarrow \infty$.

To obtain the asymptotic form of $\hat{V}_x[ \{ \phi \}, l_i]
\phi_{i}/ \phi_{i}$ we start from the asymptotic behaviour of the
atomic HF orbitals.  Handy \textit{et al.}\cite{handy69} have shown 
that the asymptotic form of the atomic HF orbitals as 
$r\rightarrow\infty$ is dominated by an exponential decay with 
exponent $\alpha = (-2\epsilon_{HO})^{\frac{1}{2}}$, where $HO$ 
denotes the highest occupied orbital. 
Handler \textit{et al.}\cite{handler80} derived the full asymptotic 
form,
\begin{equation}
\phi_{i}(r) \sim r^{\beta_{i}+1} ( a_{i} + b_{i}r^{-1} + \ldots )
e^{-\alpha r},
\label{eq:2.3}
\end{equation}
where we use $\sim$ to denote ``asymptotically approaches as
$r\rightarrow \infty$''.  They found expressions for $\beta_i$ which
can be summarised as
\begin{equation}
 \begin{array}{lll}
      i   &       l_i            & \beta_i              \\ \hline
   =   HO &  =    l_{HO}         & \beta                \\
  \neq HO &  \neq l_{HO}         & \beta-|l_i-l_{HO}|-1 \\
  \neq HO &  =    l_{HO} \neq 0  & \beta-3              \\
  \neq HO &  =    l_{HO}=0       & \beta-2(l_{min}+1)   \\
 \end{array}
\label{eq:2.4}
\end{equation}
where $\beta = (Z-(2N-1)-\alpha)/\alpha$, there are $2N$ electrons in
the atomic configuration, and $l_{min}$ is the lowest non-zero $l$
value in the configuration.

In Eq.~(\ref{eq:2.4}) and the rest of this section we limit our
analysis to closed shell atoms to aid clarity.  The generalisation to
open shell atoms using spin averaging or eigenfunctions of the total
angular momentum is straightforward but algebraically complex, and
does not change our conclusions.

Eqs.~(\ref{eq:2.3}) and (\ref{eq:2.4}) do not apply to
$s$-electron-only atoms.  In this case the exponent for each orbital
is different and is given by $\alpha_i =
(-2\epsilon_{i})^{\frac{1}{2}}$, and it is straightforward to
demonstrate that the sum of the exchange terms in Eq.~(\ref{eq:2.2})
approaches zero exponentially as $r$ increases.  We do not consider
these pure-$s$ atoms in what follows.

Expressing the HF exchange potential in spherical polar co-ordinates 
and taking the limit as $r\rightarrow \infty$ gives
\begin{eqnarray}
 \hat V_x[ \{ \phi \},l_i ] \phi_i(r) & \sim & - \sum_j \sum_l
 (2 l_j +1) w_j \left(
  \begin{array}{lll}
    l_j & l & l_i \\
    0   & 0 & 0   
  \end{array}
  \right)^2 \nonumber \\*
& & \times \frac{\phi_j}{r^{l+1}}
 \int_0^{\infty} r^l \phi_j \phi_i dr\;,
\label{eq:2.5}
\end{eqnarray}
where the Wigner $3j$ notation has been used for triple integrals over
spherical harmonics.  The occupation of shell $j$ is denoted by $w_j$,
where $w_j=1$ for a full shell.

We define the effective exchange potential for orbital $i$ as
\begin{equation}
V^x_i(r) =\frac{ \hat V_x[ \{ \phi \}, l_i ]  \phi_i(r) }
               { \phi_i(r)                              }.
\label{eq:2.6}
\end{equation}
Next we explicitly separate the exchange interaction between electrons 
within shell $i$ from the remaining exchange, and use the asymptotic 
forms for the orbitals of Eq.~(\ref{eq:2.3}), to obtain
\begin{widetext}
\begin{eqnarray}
V^x_i(r)  & \sim &
               - \frac{1}{r}
               - \sum_{k=1}^{l_i}
               (2 l_i +1) w_i \left(
               \begin{array}{lll}
               l_i & 2k & l_i \\
               0   & 0 & 0   
               \end{array}
               \right)^2
               \frac{1}{r^{2k+1}}
               \int_0^{\infty} r^{2k} \phi_i \phi_i dr  \nonumber \\*
    & &        -\sum_{j \neq i} \sum_l
               (2 l_j +1) w_j \left(
               \begin{array}{lll}
               l_j & l & l_i \\
               0   & 0 & 0   
               \end{array}
               \right)^2
               r^{n_{ij}}
               \left(
               a_{ij}+b_{ij}r^{-1}+\ldots
               \right)
               \int_0^{\infty} r^l \phi_j \phi_i dr \;,
\label{eq:2.7}
\end{eqnarray}
\end{widetext}
where $n_{ij}=\beta_j-\beta_i-l-1$.  The first term in
Eq.~(\ref{eq:2.7}) is always present and is a part of the self-interaction 
correction (SIC) which is the same for all orbitals. 
The second term is part SIC and part exchange interaction, 
and is non-zero only for $l_i > 0$. If present this introduces terms
$\propto 1/r^{2k+1}$ for $k=1,\ldots,l_i$ which are different for each
orbital.

The third term is the remainder of the exchange interaction, and we
consider the maximum power $n_{ij}$ that appears in this expression
with a non-zero coefficient.  Given that the Wigner $3j$ symbol is
non-zero only for $|l_i-l_j|\leq l \leq l_i+l_j$ and $l_i+l_j+l$ even,
and that the integral is zero for $l=0$, we find that the maximum
$n_{ij}=n_i$ takes the values
\begin{equation}
 \begin{array}{llrrr}
      i   &       l_i            & n_i     & l            \\ \hline
   =   HO &  =    l_{HO}         & -(2m+2) & m            \\
  \neq HO &  \neq l_{HO}         & 0       & |l_i-l_{HO}| \\
  \neq HO &  =    l_{HO} \neq 0  & 0       & 2            \\
  \neq HO &  =    l_{HO}=0       & 0       & l_{min}      \\
 \end{array}
\label{eq:2.8}
\end{equation}
where $m=\min_{l_i \neq l_{HO}} |l_i-l_{HO}|$ and $l_{min}$ is 
as defined for Eq.~(\ref{eq:2.4}).

Using Eq.~(\ref{eq:2.8}) we may write the effective exchange potential
in the limit $r \rightarrow \infty$ as
\begin{eqnarray}
V^x_{i}(r)    &  \sim & -\frac{1}{r}
                - \sum_{\{lj\}|n_{ij}=0}
                  w_j
                  \left(a_{ijl}+b_{ijl}r^{-1}\right)
                  \int_0^{\infty} r^l \phi_j \phi_i dr \nonumber \\*
& &             - \sum_{\{lj\}|n_{ij}=-1}
                  w_j
                  a_{ijl}r^{-1}
                  \int_0^{\infty} r^l \phi_j \phi_i dr 
                + \mathcal{O}(r^{-2}) \;,              \nonumber \\*
\label{eq:2.9}
\end{eqnarray}
where the Wigner $3j$ and its pre-factor in Eq.~(\ref{eq:2.7}) have
been subsumed into the parameters $a_{ijl}$ and $b_{ijl}$, and the
sums are taken over the $l$ and $j$ that give $n_{ij}=0$ and
$n_{ij}=-1$.  For $i=HO$, only the SIC term appears since
$a_{ijl}=b_{ijl}=0$.  Note that for $i \neq HO$, $V^x_i(\infty)$ is
non-zero and the effective exchange potential contains a $r^{-1}$ term
in addition to the SIC.

Using the same line of argument, the pseudo-atom effective exchange
potential at large distances is given by
\begin{eqnarray}
\tilde{V}^x_{i}(r)          &  \sim & - \frac{1}{r}
                - \sum_{\{lj\}|n_{ij}=0}
                  \tilde{w}_j
                  \left(a_{ijl}+b_{ijl}r^{-1}\right)
                  \int_0^{\infty} r^l \tilde{\phi}_j \tilde{\phi}_i dr 
\nonumber \\*
& &             - \sum_{\{lj\}|n_{ij}=-1}
                  \tilde{w}_j
                  a_{ijl}r^{-1}
                  \int_0^{\infty} r^l \tilde{\phi}_j \tilde{\phi}_i dr 
                + \mathcal{O}(r^{-2})\;,
\nonumber \\*
\label{eq:2.10}
\end{eqnarray}
again with $a_{ijl}=b_{ijl}=0$ for $i=HO$.  The sum over $l,j$ is as
before, and a similar non-zero limit and $r^{-1}$ term appear.  The
occupation of pseudo-shell $j$, $\tilde{w}_j$, is defined as for the
AE atom, but is zero for core states.  Since the integrals in
Eq.~(\ref{eq:2.9}) and Eq.~(\ref{eq:2.10}) are not in general equal,
the limits for the AE and pseudo-atom effective exchange potentials
are not equal and do not cancel in Eq.~(\ref{eq:2.2}).  In addition,
exchange terms due to core-valence interactions will not cancel as no
core orbitals are present in the pseudo-atom.

The asymptotic behaviour of the pseudopotential can be obtained to
$\mathcal{O}(r^{-2})$ from the explicit forms of the external,
Hartree, and exchange potentials.  Starting with the AE effective
potential, Eq.~(\ref{eq:1.3}) can be expressed as \newline
\begin{eqnarray}
  V^{eff}_i & \sim & -\left( Z - (2N-1) \right) r^{-1} \nonumber \\*
            & & - \sum_{\{lj\}|n_{ij}=0}
                  w_j
                  \left(a_{ijl}+b_{ijl}r^{-1}\right)
                  \int_0^{\infty} r^l \phi_j \phi_i dr \nonumber \\*
            & & - \sum_{\{lj\}|n_{ij}=-1}
                  w_j
                  a_{ijl}r^{-1}
                  \int_0^{\infty} r^l \phi_j \phi_i dr
                + \mathcal{O}(r^{-2})\;.               \nonumber \\*
\label{eq:2.11}
\end{eqnarray}
If we write the asymptotic form of the pseudopotential as
\begin{equation}
\tilde{V}_i \sim s_i - (Z_{val}+g_i) r^{-1}+ \mathcal{O}(r^{-2}),
\label{eq:2.12}
\end{equation}
where $s_i$ and $g_i$ are constants then the equivalent expression 
for the pseudo-atom effective potential is
\begin{eqnarray}
\tilde{V}^{eff}_i & \sim & s_i - \left( 
                  Z_{val} - (2N_{val}-1) + g_i 
                    \right) r^{-1}
 \nonumber \\*
            & & - \sum_{\{lj\}|n_{ij}=0}
                  \tilde{w}_j
                  \left(a_{ijl}+b_{ijl}r^{-1}\right)
                  \int_0^{\infty} r^l \tilde{\phi}_j \tilde{\phi}_i dr 
 \nonumber \\*
            & & - \sum_{\{lj\}|n_{ij}=-1}
                  \tilde{w}_j
                  a_{ijl}r^{-1}
                  \int_0^{\infty} r^l \tilde{\phi}_j \tilde{\phi}_i dr 
                + \mathcal{O}(r^{-2})\;, \nonumber \\*
\label{eq:2.13}
\end{eqnarray}
where there are $2N_{val}$ valence electrons.  Since $V^{eff}_i =
\tilde{V}^{eff}_i$ for $r>r_{ci}$ we may equate the coefficients of
$r^{-1}$ and $r^{0}$ to give
\begin{widetext}
\begin{equation}
s_i =
\left\{
\begin{array}{ll}
   0  & i=HO \\
   \sum_{\{lj\}|n_{ij}=0}
   a_{ijl} \left[
   \tilde{w}_j \int_0^{\infty} r^l \tilde{\phi}_j \tilde{\phi}_i dr-
   w_j         \int_0^{\infty} r^l        \phi_j         \phi_i  dr
   \right] & i \neq HO\;, \\
\end{array}
 \right.
\label{eq:2.14}
\end{equation}
and, assuming that the charge of the AE and pseudo-atoms are equal
($Z-2N = Z_{val}-2N_{val}$),
\begin{equation}
g_i =
\left\{
\begin{array}{lll}
   0  & & i=HO \\
   - & \left(
   \sum_{\{lj\}|n_{ij}=0}
   b_{ijl} \left[
    \tilde{w}_j \int_0^{\infty} r^l \tilde{\phi}_j \tilde{\phi}_i dr-
    w_j         \int_0^{\infty} r^l        \phi_j         \phi_i  dr
   \right]  \right. + & \\
    & \left.~ 
   \sum_{\{lj\}|n_{ij}=-1}
   a_{ijl} \left[
    \tilde{w}_j \int_0^{\infty} r^l \tilde{\phi}_j \tilde{\phi}_i dr-
    w_j         \int_0^{\infty} r^l        \phi_j         \phi_i  dr
    \right] \right) & i \neq HO\;. \\
\end{array}
 \right.
\label{eq:2.15}
\end{equation}
\end{widetext}

For $i\neq HO$, $s_i$ and $g_i$ are generally non-zero since the AE
and pseudo-orbitals differ in the core region.  In addition, terms due
to exchange between state $i$ and core states have no counterpart in
the pseudo-atom to cancel with, as $w_j=1$ and $\tilde{w}_j=0$ for the
core states.  This orbital dependent ``offset potential'' and ``ghost
charge'' result from a remnant of the AE exchange interaction remaining
in the pseudopotential as it is not completely removed by the exchange
interaction in the pseudo-atom, and is generally due to both
valence-valence and valence-core exchange.  Because these terms are
orbital dependent the pseudopotential itself is non-local over all
space and does not satisfy the locality criteria of Eq.~(\ref{eq:2.1})
for any $r_{loc}$.  It should be stressed that this extreme
non-locality is necessary if we require our HF pseudopotential to be
both norm conserving and to reproduce the AE valence eigenvalues and
orbitals outside of the core region.

To illustrate the physical consequences of this behaviour we consider
the total energy of a collection of identical pseudo-atoms.  The total
energy should tend towards the total energy of the isolated
pseudo-atoms as they are separated further apart.  This is true if
each channel of the pseudopotential approaches $-Z_{val}/r$ for large
$r$. For the extreme non-local behaviour found here this is not the
case, and an unphysical interaction persists at large distances. For
extended systems this is catastrophic because the total energy is
undefined.

Extreme non-locality may be avoided in the construction of a
pseudopotential by using a different configuration to construct each
channel, such that the configuration used to construct channel $l$ 
has $l=l_{HO}$.  Although this is practical for H and He, for heavier
atoms this requires the construction of pseudopotentials from
electronic configurations far from those we wish to use them for.  It
would not be reasonable to expect the resulting pseudopotentials to be
transferable to systems of interest, and they would not reproduce the 
valence states of the neutral isolated atom.

\subsection{The Kleinman-Bylander form}
\label{subsec:KB}

The Kleinman-Bylander form is expressed in terms of operators that
project the orbitals onto a set of basis functions, and takes the
general form
\begin{equation}
\hat{V}^{KB}=\tilde{V}_{local}(r) + \sum_{ij} | \tilde{\psi}_i \rangle D_{ij}
\langle \tilde{\psi}_j |\;,
\label{eq:2.1.1}
\end{equation}
for some $\{\tilde{\psi}\}$.
The pseudopotentials discussed in section \ref{sec:def_pp} may be
expressed in this form if we define
\begin{equation}
D_{ij}=\langle \tilde{\psi}_i| \tilde{V}_j - \tilde{V}_{local} |
\tilde{\psi}_j \rangle,
\label{eq:2.1.2}
\end{equation}
where the $\tilde{\psi}_i$ are the eigenstates of the pseudo-atom.  If
a complete set of $\tilde{\psi}_i$ is included in the sum in
Eq.~(\ref{eq:2.1.1}), this form would be exactly equal to
Eq.~(\ref{eq:1.7}).  In practice a more useful approach is to use a
small number of localised, bound, atomic states (such as those used to
construct the original pseudopotential) with $\tilde{V}_{local}(r)$
chosen such that $\tilde{V}_{local} \rightarrow -Z_{val}/r$ as $r
\rightarrow \infty$.  For this choice of localised bound states,
$\hat{V}^{KB}$ is well-behaved at large $r$, since the subspace
covered by the projection does not include this asymptotic region -
the non-local part of the pseudopotential has been localised by
representing it within a localised subspace.

We conclude that converting a norm-conserving HF pseudopotential to
Kleinman-Bylander form removes the long ranged non-locality, and so
allows these potentials to be implemented computationally in a manner
which avoids the unphysical behaviour described previously.  It should
be noted that the ``frozen remnant'' of the atomic exchange interaction
which makes up the non-Coulombic tail is still present (although
limited to a subspace represented by the pseudo-orbitals), and it may
still lead to significant errors.

\subsection{DFT pseudopotentials}
\label{subsec:DFT}

Extreme non-locality does not occur for pseudopotentials constructed
within Kohn-Sham DFT because the effective potential that represents
the electron-electron interaction is the same for all electrons (i.e.,
it is local).  More explicitly, in DFT the local exchange-correlation
potential takes the place of the exchange potential in
Eq.~(\ref{eq:2.2}) and hence the equivalent expression for the
pseudopotential is
\begin{equation}
 \tilde{V}_{i}(r) =
- \frac{Z}{r}
+ V_h[ \rho - \tilde{\rho} ]
+ {V}_{xc}[        \rho  ] 
- {V}_{xc}[ \tilde{\rho} ], \;\;\;\; r>r_{ci}.
\label{eq:2.2.1}
\end{equation}
For $r> r_c$, where $r_c =\max \left[r_{ci}\right]$, all 
$\tilde{V}_{i}(r)$ are the same, so the pseudopotential is 
local outside of the core region.

Within the Optimised Potential Method (OPM) \cite{talman76,kummel03}
the exchange-correlation energy functional is replaced by an ``exact''
exchange energy functional (possibly with an added approximate
correlation functional) defined explicitly in terms of the Kohn-Sham
orbitals and implicitly in terms of the charge density.  This exact
exchange (EXX) functional takes the form of the Fock term of HF, but
is evaluated using the KS orbitals and hence the calculation remains
KS-DFT \cite{nl-ext} and the OPM exchange energy does not equal the HF
exchange energy.  Norm-conserving pseudopotentials have been defined
within this theoretical picture by Engel \emph{et al.} and others, and
``spurious long range structure''\cite{hock98,moukara00,engel01}
consistently appears in these pseudopotentials.

Given the highly non-local nature of the Fock term it seems reasonable
that the exchange potential is also an extremely non-local functional 
of the density.  We also expect the exchange potential to
approach zero as $r\rightarrow \infty$, since this behaviour is part
of the definition of the EXX potential itself \cite{talman76}.  These
two observations, together with Eq.~(\ref{eq:2.2.1}), explain the
existence of the structure found by Engel \emph{et al.}~\cite{engel01}
They found that pseudopotentials constructed by removing this 
long range structure resulted in improved 
bond distances and energies for simple dimers and bulk Al.
Note that the OPM case is in some regards similar to 
that found in the HF case, but it is essentially different in 
that the potential is local for $r>r_{c}$ and approaches 
$-Z_{val}/r$ faster than $1/r$.

For local (e.g., the local density approximation or LDA) or 
semi-local (e.g., the generalised gradient approximation or GGA) 
exchange-correlation functionals we note that $\rho$ and 
$\tilde{\rho}$ differ by the exponentially decaying core 
contribution and obtain the relation
\begin{equation}
\lim_{r \rightarrow \infty} \tilde{V}_{i}(r) = - \frac{Z_{val}}{r} +
\mathcal{O}(e^{-\alpha r}).
\label{eq:2.2.2}
\end{equation}
The absence of an offset or ghost charge is a consequence of the
locality of the approximate exchange-correlation functional. It is
clear that there is no DFT analogue of the extreme non-locality
present in HF pseudopotentials.

\section{Results}
\label{sec:results}

\subsection{Size of the extreme non-local terms}
To construct a pseudopotential from an AE atomic HF calculation we
must define a form for the pseudo-orbitals inside the core radius,
$f_i$.  We chose the Troullier-Martins~\cite{troullier91} form
\begin{equation}
f_i(r) = r^{l+1}\exp \left[ \sum_{m=0}^{6} c_m r^{2m} \right],
\label{eq:3.1}
\end{equation}
which is based on the observation that the pseudo-orbitals are
smoother if the odd derivatives of the screened pseudopotential at the
nucleus are set to zero.  The coefficients are determined by requiring
norm-conservation of the charge in the core region, continuity of
$\tilde{\phi}_i$ and its first four derivatives at $r_{ci}$, and the
requirement that $\tilde{V}^{eff}_i$ has zero curvature at the origin.

As an example we consider Ne ($1s^22s^22p^6$) and perform a numerical
HF calculation and construct $\tilde{\phi}$ for the $2s$ and $2p$
orbitals.  We then obtain $s$ and $p$ pseudopotentials as described in
section \ref{sec:def_pp}.  We also calculate the $r^{-1}$ and $r^{0}$
contributions to the asymptotic behaviour directly from the
pseudo-orbitals.

First we note that, for Ne, the highest occupied orbital is
$i=2p$. From Eqs.~(\ref{eq:2.7}) and Eq.~(\ref{eq:2.8}) it follows 
that the $p$ part of the pseudopotential approaches $-Z_{val}/r$ 
as $r^{-3}$ due to the $2p-2p$ exchange interaction.

For the $s$ part, the exchange interaction contributes the $2s-2p$
term, which is the only interaction appearing in
Eqs.~(\ref{eq:2.11}) and (\ref{eq:2.13}).  To obtain values for
$s_{2s}$ and $g_{2s}$ we first obtained the asymptotic behaviour of
the effective potential $V^{eff}_i(r)$ by direct substitution of the
asymptotic form of the orbital into Eq.~(\ref{eq:1.2}).  Equating
powers of $r$ leads to
\begin{equation}
(\epsilon_{2s}-\epsilon_{2p}) = -a_{2p,2s,1} \int_0^{\infty} r \phi_{2p}
\phi_{2s} dr \;,
\label{eq:3.2}
\end{equation}
and
\begin{equation}
2\sqrt{2}|\epsilon_{2p}|^{\frac{1}{2}} = 
     b_{2p,2s,1} \int_0^{\infty} r \phi_{2p} \phi_{2s} dr\;.
\label{eq:3.3}
\end{equation}

Using the numerical values of the integrals and eigenvalues we obtain
the values of $a_{2p,2s,1}$ and $b_{2p,2s,1}$, and by using
Eqs.~(\ref{eq:2.14}) and (\ref{eq:2.15}) we obtain $s_{2s} = 0.0095$
a.u.~(0.26~eV) and $g_{2s}=0.023$ \textit{e}, that is, a positive
offset to the potential and a positive ghost charge.

\begin{figure}
\includegraphics{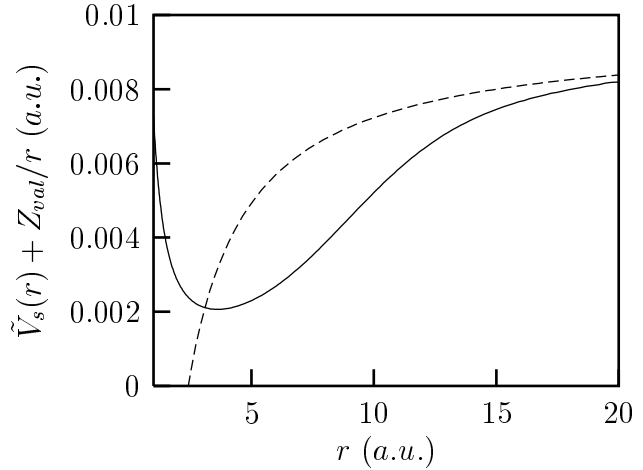}
\caption{\label{fig1} The difference between the $s$ part of the Ne
pseudopotential and the Coulomb potential, $\tilde{V}_{s} +
Z_{val}/r$, against radial distance (solid line), and the calculated
asymptotic form (dashed line).}
\end{figure}

The difference between the $s$ part of the Ne pseudopotential and the
Coulomb potential, $\tilde{V}_{s} + Z_{val}/r$ is shown in
Fig.~\ref{fig1}. It is clear that the pseudopotential does not
approach the Coulomb potential for large $r$.  The calculated
asymptotic form, $s_{2s} + g_{2s}/r$, is also shown.  The difference
between the pseudopotential and the calculated asymptotic form at
large $r$ arises mainly from a $1/r^4$ term which was not included in
the asymptotic form.

The magnitude of the deviation from the Coulomb potential at large $r$
shown in Fig.~\ref{fig1} is typical of that found for all $p$ and
$d$-block atoms. We conclude that the non-locality persists far from
the nucleus and is of significant magnitude for these atoms.  For
$s$-block atoms the $s$ channel is due to the $HO$ orbital, hence
$s_{HO}=g_{HO}=0$. The $p$ and $d$ channels are generally obtained
from excited atomic states for which the offset is typically $\sim
10^{-5}$~a.u. for the channel which does not correspond to the $HO$
orbital.

\subsection{Localising the pseudopotential}

Here we consider several methods for removing the non-Coulombic tails
of the pseudopotentials in order to localise them within some radius
$r_{loc}$. In what follows we define the localisation radius,
$r_{loc}$, as in Eq.~(\ref{eq:2.1}), with $\delta=5 \times 10^{-6}$
a.u.

Previous workers (for example, Greef and Lester~\cite{greeff98}) 
have expanded pseudopotentials in a Gaussian basis, a standard 
form for quantum chemistry applications.
Due to the functional form of this expansion the parameterised 
pseudopotential is non-local in a finite region surrounding the 
nucleus, and approaches the Coulombic form as $r\rightarrow \infty$, 
hence the expansion may be used as a localisation procedure.
Generally this expansion is obtained numerically by requiring 
the parameterised pseudopotential to reproduce the
pseudo-states of the original pseudopotential to within a given 
accuracy.~\cite{barthelat77}
Although this method can work well it is difficult to control the 
smoothness and the localisation radius, $r_{loc}$, since the 
error in the expanded pseudopotential (and accompanying 
pseudo-states) must compensate for the absence of the long range 
behaviour present in the original pseudopotential.
We prefer to control the localisation of the original pseudopotential 
and then, if desired, expand the resulting potential in a 
Gaussian basis.
Gaussian expansions are not considered further in this paper, but
we will present parameterised pseudopotentials obtained in this way 
in a separate paper\cite{trail04}.

We have freedom in our choice of $r_{ci}$ and $f_i(r)$, and by varying
these quantities we might be able to localise the pseudopotential in
the sense of Eq.~(\ref{eq:2.1}), since variations in these quantities
will alter the integrals in Eqs.~(\ref{eq:2.14}) and (\ref{eq:2.15}).
For Ne (and for most atoms) the first term in Eqs.~(\ref{eq:2.14}) and
(\ref{eq:2.15}) is larger than the second because the AE orbitals
possess radial intervals where they are negative, while the
pseudo-orbitals are positive for all $r$.  In order to decrease the
value of this integral, causing $s_{2s}$ and $g_{2s}$ to become closer
to zero, $f_i(r)$ might be chosen such that, on average, the electron
is closer to the nucleus. This might be achieved by increasing
$r_{ci}$, allowing the maximum of $\tilde{\phi}_{2s}$ to be closer to
the nucleus.  Also, the form of $f_i(r)$ may be altered (a convenient
way of achieving this is to relax the continuity of the $3^{rd}$
and/or $4^{th}$ derivatives).  Calculations suggest that neither of
these approaches is useful - for a good cancellation of the integrals
in Eqs.~(\ref{eq:2.14},\ref{eq:2.15}) $r_{ci}$ must be very large, or
$f_i(r)$ must take a form such that the corresponding pseudopotential
is far from smooth.

Next we consider an elegant procedure described by 
Engel~\emph{et al.}\cite{engel01} and used by them to remove the 
(local) intermediate range structure present in OPM pseudopotentials.
This approach employs a self-consistent modification of the 
pseudo-orbitals over all space, while conserving the norm. The same 
method may be applied to localise HF pseudopotentials.

The pseudopotential at iteration $n$ is denoted by $\tilde{V}_{i;n}$,
and the associated pseudo-orbitals are $\{ \tilde{\phi}_{i} \}_n$.
The pseudopotentials calculated in section~\ref{sec:def_pp} correspond
to $n=1$. We construct an effective potential which is not yet 
self-consistent,
\begin{equation}
 \tilde{V}^{eff}_{i;n} =
- \frac{Z}{r} + V_h[ \rho_{core}]
+ V_h[ \tilde{\rho}_n ] 
+ \frac{ \hat{V}_x[ \{ \tilde{\phi} \}_n, l_i]
     \tilde{\phi}_{i;n} }{ \tilde{\phi}_{i;n} }\;,
\label{eq:3.10}
\end{equation}
where $-Z/r + V_h[ \rho_{core} ]$ is the Coulomb$+$Hartree potential 
due to the ionic core.
Equation (\ref{eq:3.10}) is the external potential we desire 
outside of the core radius together with the effective interaction 
potential of the pseudo-orbitals $\{ \tilde{\phi}_{i} \}_n$.
Taking the correct asymptotic form for the orbitals, we integrate 
in from $\infty$ to $r_{ci}$ at energy $\epsilon_i$ to obtain 
the new orbitals 
$\{ \tilde{\phi}_{i} \}_{n+1}$ in this region, and these are 
normalised by requiring that the norm of the newly generated and original 
pseudo-orbitals outside of $r_{ci}$ are equal.
In the core region the new orbital is constructed by 
using the Troullier-Martins form and criteria described at 
the beginning of this section.
Taking $\{ \tilde{\phi}_{i} \}_{n+1}$, we invert the HF equations as 
described in section \ref{sec:def_pp} to find the new external potential 
that results in these orbitals $\{ \tilde{\phi}_i \}_{n+1}$ on 
solution of the HF equations, $\tilde{V}_{i;n+1}$.

The new set of orbitals are then used in Eq.~(\ref{eq:3.10}) to generate the 
effective potential $\tilde{V}^{eff}_{i;n+1}$ and the entire process repeated 
until self-consistency is achieved - we use 
$|\tilde{V}_{i;n+1} - \tilde{V}_{i;n}| < 10^{-6}$  a.u. for all $r$.
At self-consistency $\tilde{V}_{i;n}$ is the pseudopotential we require.
This method finds the HF orbitals that take the Troullier-Martins 
form in the core region, result from a local ionic potential outside 
of the core region, and have the same eigenvalues and norm as the 
original AE atom valence states.

A number of difficulties presented themselves with this method.
We found that for atoms with few valence  electrons it was not 
possible to achieve self-consistency.
This appeared to be because the analytic form 
used for the orbitals in the core region was not general enough 
to give a self-consistent solution.
Furthermore, although Engel's approach conserves the norm of the 
original AE states, it does not conserve the magnitude  
or the radial derivative of the orbitals at $r_{ci}$.
Consequently the new pseudopotential does not conserve the logarithmic 
derivative or its energy derivative, and does not reproduce the 
scattering properties of the AE atom.
For example, for the $s$-channel of Ne the new pseudopotentials
results in a $-0.24\%$ change in the 
magnitude of the orbital at the core radius, a $+3.9\%$ change 
in the logarithmic derivative, and a $+0.40\%$ change in the 
energy derivative of the logarithmic derivative.
These errors may be acceptable, although the error in the 
logarithmic derivative might be problematic.

By introducing an interface region between the core 
region and the valence region (where the pseudopotential is ionic) it 
is in principle possible to modify Engel's method such that the magnitude 
of the orbital, the logarithmic derivative and the energy derivative of 
the logarithmic derivative are all conserved at the core radius.
This was attempted using the Troullier-Martins form together with 
additional criteria.
We found that either self-consistency could not be achieved or that 
the resulting pseudopotentials were far from smooth.

The method we chose to use is to apply a transformation
to the original pseudopotential such that deviations from the 
Hartree$+$Coulombic ionic core potential are removed beyond a 
certain radius but desirable properties of the original 
potentials are preserved.
We investigated the transformation 
\begin{equation}
\tilde{V}^{loc}_i(r) =
  \left\{
  \begin{array}{ll}
    \gamma_i(r) + \tilde{V}_i(r)         & r < R_i \vspace{+0.2cm} \\
    e^{-\alpha(r-R_i)^2}                 &         \\
    ~ \times \left( 
       \gamma_i(r) + \tilde{V}_i(r)
       - V_h[ \rho_{core} ] + \frac{Z}{r}
    \right)                              & \\
    ~ + V_h[ \rho_{core}] - \frac{Z}{r}  & r \geq R_i \;,
  \end{array}
  \right.
\label{eq:3.4}
\end{equation}
where $-Z/r + V_h[ \rho_{core} ]$ is the ionic potential.
The Gaussian decay of the non-Coulombic part of the potential
occurs over a distance $\sim 1/\alpha^{-\frac{1}{2}}$, and 
the zeroth and first derivatives of $\tilde{V}^{loc}_i$ are 
continuous at $R_i$.
The precise value of $\alpha$ has very little influence on the final 
result - the cutoff radius, $R_i$, governs the quality of the 
pseudopotential.

For the simple cutoff function corresponding to $\gamma_i=0$ and 
$\alpha^{-\frac{1}{2}} = 0.15$ a.u. we 
chose $R_i$ as the minimum value for which the 
$|\epsilon_i - \tilde{\epsilon}^{loc}_i| \le 10^{-5}$ a.u., 
where the $\tilde{\epsilon}^{loc}_i$ are the eigenvalues 
of the new pseudopotential.
For Ne this new pseudopotential is local (in the sense of
Eq.~(\ref{eq:2.1})) outside of a radius of $r_{loc}=3.35$~a.u.
Although this is large in comparison with the core radii of the 
orbitals themselves, it may be acceptable, although it would be 
expensive for DMC.  Moreover,
this approach enforces a Gaussian decay on a monotonically increasing
function, essentially creating a (smoothed) step-like behaviour in the
pseudopotential, and it seems likely that this would introduce
transferability problems.  Note that if we seek to remove this
step-like behaviour by decreasing $\alpha$ then $r_{loc}$ rapidly
becomes impractically large.

A more successful approach is to choose $R_i=r_{ci}$ and 
$\alpha^{-\frac{1}{2}} = r_c/16$ (as before, the final result is 
insensitive to the choice of this parameter) 
so that the pseudopotential is non-local in a small region and 
no step-like behaviour occurs.
In order to reproduce the eigenvalues of the original pseudopotential, 
$\gamma_i$ must be non-zero, and we use
\begin{equation}
\gamma_i(r) =
  \left\{
  \begin{array}{ll}
    q_i + p_i r^4 \left(1 - \frac{2}{3R_i^2} r^2 \right)  & r <    R_i \\
    q_i + p_i \frac{R_i^4}{3}                             & r \geq R_i \;,
  \end{array}
  \right.
\label{eq:3.5}
\end{equation}
where $\alpha$ is chosen as before and $q_i$ and $p_i$ are 
orbital dependent parameters. This function is continuous and smooth
at $r=R_i$, and the second derivative is zero at $r=0$ (see the 
discussion following Eq.~(\ref{eq:3.1})).
Our goal is to search for values of these parameters such that the 
localised pseudopotential preserves certain desirable features of 
the original pseudopotential. We require that the conditions
\begin{eqnarray}
\left[
 \frac{\partial}{\partial r} \ln \tilde{\phi}^{loc}_i -
 \frac{\partial}{\partial r} \ln \tilde{\phi}^{~}_i
\right]_{R_i}=0 \;, \nonumber\\*
\tilde{\epsilon}^{loc}_i-{\epsilon}^{~}_i=0 \;,
\label{eq:3.6}
\end{eqnarray}
are satisfied in order to reproduce the original eigenvalues and 
aid transferability. 
Conservation of these two quantities takes precedence over 
norm conservation and we do not require the norm of 
$\tilde{\phi}^{loc}_i$ to be equal to the norm of 
$\tilde{\phi}^{~}_i$ as we could not find a reliable method to achieve 
this and give a smooth pseudopotential.

Conservation of the logarithmic derivative of each pseudo-orbital at 
$R_i$ may be achieved exactly, but a simpler and more convenient 
approximation is used here.
Starting with the equations for the original pseudo-orbital and the
pseudo-orbital resulting from the localised pseudopotential we may
follow a similar derivation to that used to arrive at the standard 
norm-conservation condition to obtain
\begin{eqnarray}
\frac{\tilde{\phi}^{loc}_i \tilde{\phi}^{~}_i}{2} 
\left[
\frac{\partial}{\partial r} \ln \tilde{\phi}^{loc}_i -
\frac{\partial}{\partial r} \ln \tilde{\phi}^{~}_i
\right]_{R_i}&=& \nonumber \\*
  \int_0^{R_i} \tilde{\phi}^{loc}_i
                  \Delta V
                  \tilde{\phi}^{~}_i dr 
  - (\tilde{\epsilon}^{loc}_i&-&{\epsilon}_i)
 \int_0^{R_i} \tilde{\phi}^{loc}_i 
             \tilde{\phi}^{~}_i dr \;, \nonumber \\*
\label{eq:3.7}
\end{eqnarray}
where the eigenvalues are not yet constrained to be equal and 
$\Delta V = \tilde{V}^{eff,loc}_{i}-\tilde{V}^{eff}_{i}$ is the 
difference between the effective potentials for calculations 
carried out with the original and localised pseudopotentials (for 
the configuration used to construct the pseudopotentials).
Our approximation is to take $\Delta V$ to be close to the difference 
between the external potentials, 
$\Delta V \approx \tilde{V}^{loc}_{i}-\tilde{V}_{i} = \gamma_i$, 
for $r<R_i$.

We start by using simple iteration. For a fixed $q_i$ we take an initial 
value of $p_i=0$ and solve for the resulting pseudo-orbitals. We then 
update $p_i$ by solving $\int_0^{R_i} \tilde{\phi}^{loc}_i \gamma_i 
\tilde{\phi}^{~}_i dr =0$ for $p_i$ and continue until self-consistency 
is achieved.
This is repeated for different values of $q_i$ and the condition 
$\tilde{\epsilon} ^{loc}_i-{\epsilon}^{~}_i=0$ is enforced by 
straightforward bisection.
This process is repeated to obtain self-consistency across all 
channels, and we obtain eigenvalues which differ by $< 10^{-7}$~a.u., 
and $\int_0^{R_i} \tilde{\phi}^{loc}_i \gamma_i \tilde{\phi}^{~}_i dr =0$ 
to numerical precision.
Note that as we have not taken into account the change in the 
effective potential due to changes in the exchange and Hartree 
potentials, the LHS of Eq.~(\ref{eq:3.7}) is not zero, and there 
will be a small change in the logarithmic derivative.

\begin{figure}
\includegraphics{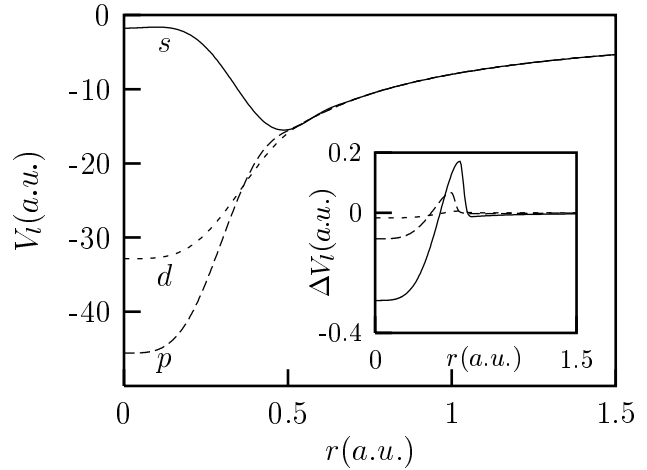}
\caption{\label{fig2} The localised pseudopotential for Ne.
The inset shows the difference
between the localised and original pseudopotentials.}
\end{figure}

Localised pseudopotentials for Ne are shown in Fig.~\ref{fig2}.
The $s$ and $p$ channels were obtained from the neutral ground state,
while the $d$ part was obtained from an excited ionic state
($1s^{2}2s^{1}2p^{2.75}3d^{0.25}$). All excited state 
configurations used to obtain
pseudopotential channels not bound in the ground state
configuration are taken from Bachelet \emph{et al.}\cite{bachelet82}
Inset in the same figure is the difference between the localised
pseudopotential and the original numerically exact pseudopotential.
This figure demonstrates that our localisation procedure deforms the 
pseudopotentials by a small amount, mostly within the core region.

\begin{figure}
\includegraphics{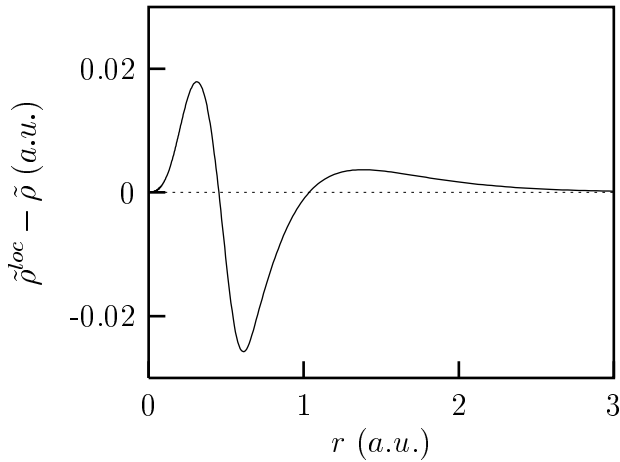}
\caption{\label{fig3} Difference between the charge density resulting
from the localised ($\tilde{\rho}^{loc}$) and original
($\tilde{\rho}$) pseudopotentials for Ne.  To achieve localisation a
small amount of charge near the nucleus is redistributed.  }
\end{figure}

For all atoms considered we found that this process results in a 
localised pseudopotential which modifies the form of the
associated orbitals over all space, but by a very small amount - about
$0.1\%$ of each electron is removed from a region around the core radius
and redistributed into two regions, one closer to and one further 
from the nucleus.
The removal of offset and ghost charge (and
higher order terms) results from a small change in the asymptotic
behaviour of the orbital at large $r$.  Normalisation must be
preserved, and hence the orbitals are also changed slightly at small
$r$.  This effect is illustrated in Fig.~\ref{fig3}, which shows the
difference between the radial charge density of Ne resulting from the
localised pseudopotential and the radial charge density resulting
from the original pseudopotential.

The transformed Ne pseudopotential is local (in the sense of
Eq.~(\ref{eq:2.1})) outside of a radius of $r_{loc}=0.74$~a.u.
A similarly localised pseudopotential constructed by applying 
Eq.~(\ref{eq:3.4}) with $\gamma_i=0$ (using the same $\alpha$ 
and $R_i$ values) results in a large error in the $2s$ eigenvalue 
of $-1.2\times10^{-2}$~a.u.

As the localised version of the pseudopotential is not equal to
the original pseudopotential it is not exactly norm conserving, and
does not preserve the form of orbitals outside the core region or at 
the core radius.
To illustrate the magnitude of these effects we return to Ne, 
specifically the $s$ channel as this is the major source of 
the extreme non-locality present in the original pseudopotential.
The localisation process results in a $-0.27\%$ change in the 
magnitude of the orbital at the core radius, a $-0.05\%$ change 
in the logarithmic derivative (due to assuming that the exchange 
and Hartree terms to do not contribute in Eq.~(\ref{eq:3.7})), 
and a $+0.80\%$ change in the energy derivative of the logarithmic 
derivative.

The minor relaxation of norm-conservation that accompanies the
localisation may influence the transferability of the
pseudopotential.  However, the familiar conservation of the first
order change in the radial logarithmic derivative with respect to
energy is only valid when the potential inside the core region 
is fixed.~\cite{goedecker92} 
For both the HF and DFT this is not the case even for an isolated atom, 
since $V_i^{eff}$ is a functional of $\{ \phi \}$ and each $\phi_i$ 
is a function of $\{ \epsilon \}$, and we have shown 
(from Eq.~(\ref{eq:3.7})) that the norm-conservation 
relation is replaced by
\begin{eqnarray}
 \left. \frac{\partial}{\partial \epsilon_j} \frac{\partial}{\partial
 r} \ln |\phi_i| \right|_{r_{ci}} &=& 
- \delta_{ij} \frac{2}{|\phi_i(r_{ci})|^2}
  \int_0^{r_{ci}} |\phi_i|^2 \, dr \nonumber \\*
& & + \frac{2}{|\phi_i(r_{ci})|^2}
  \int_0^{r_{ci}} |\phi_i|^2 \frac{\partial V_i^{eff}}{\partial
 \epsilon_j} \, dr. \nonumber \\*
\label{eq:3.9}
\end{eqnarray}
This applies to both the AE and pseudo-atoms.
Given that Eq.~({\ref{eq:3.9}}) tells us that first order changes 
in the radial logarithmic derivative depend on the variation of 
the effective potential with $\epsilon_i$ as well as the norm, 
and that norm-conserving pseudopotentials have been successful 
in the past, it is likely that the very small relaxation of 
norm-conservation is not significant.

\begin{table}[t]
\begin{tabular}{lrrrr} \hline \hline
Atom\ \ \ \ &
\ \ \ \ $r_{cs}$ &
\ \ \ \ $r_{cp}$  &
\ \ \ \ $r_{cd}$  &
\ \ \ \ $r_{loc}$ \\ \hline
H  & 0.50 & 0.50 & 0.50 & 0.49 \\
He & 0.60 & 0.60 & 0.60 & 0.59 \\
Li & 2.19 & 2.37 & 2.37 & 2.71 \\
Be & 1.88 & 1.96 & 1.96 & 2.20 \\
B  & 1.41 & 1.41 & 1.41 & 1.63 \\
C  & 1.10 & 1.10 & 1.10 & 1.29 \\
N  & 0.94 & 0.88 & 0.84 & 1.09 \\
O  & 0.80 & 0.75 & 0.99 & 1.13 \\
F  & 0.70 & 0.64 & 0.89 & 1.02 \\
Ne & 0.63 & 0.57 & 0.63 & 0.74 \\
Na & 2.70 & 2.85 & 2.85 & 3.25 \\
Mg & 2.38 & 2.38 & 2.38 & 2.71 \\
Al & 1.94 & 2.28 & 2.28 & 2.64 \\
Si & 1.67 & 2.01 & 2.06 & 2.36 \\
P  & 1.48 & 1.71 & 1.71 & 1.98 \\
S  & 1.33 & 1.50 & 1.50 & 1.74 \\
Cl & 1.19 & 1.34 & 1.34 & 1.55 \\
Ar & 1.09 & 1.20 & 1.31 & 1.54 \\ \hline \hline
\end{tabular}
\caption{Core radii and localisation radii (a.u.).}
\label{tab:1}
\end{table}

As a further test of the approach described above we have generated
localised pseudopotentials for H-Ar.
The core radii and localisation radii of these pseudopotentials are 
given in Table \ref{tab:1}.
Useful pseudopotentials are required to be transferable, that is,
to accurately reproduce the behaviour of the AE atom in other
environments.  In Table \ref{tab:2} we compare excitation energies of
the atoms H-Ar resulting from AE, original pseudopotentials, and 
localised pseudopotential calculations.
Excitation energies are calculated for each atom as the difference 
between the total energies of the excited and ground states for
each type of calculation.

We compare our pseudopotentials with the HF pseudopotentials of
Ovcharenko \emph{et al.}\cite{ovcharenko01}.  This is the most
appropriate comparison as both sets of pseudopotentials are soft (in
the sense of possessing no singularity at the origin).  The
pseudopotentials of Ovcharenko \emph{et al.}\cite{ovcharenko01} are,
however, parameterised, and we note that some of the differences in
accuracy will correspond to errors in the parameterisation rather
than differences in the methods of generation.

In Table \ref{tab:3} we compare the ionisation potentials 
resulting from the AE, the original pseudopotentials, the localised 
pseudopotentials, and the pseudopotentials published by Ovcharenko 
\emph{et al.}\cite{ovcharenko01} in the same manner as 
Table \ref{tab:2}. Ionisation potentials are defined for each atom 
as the difference between the total energies of the ionised and
ground states for each type of calculation.

From Table \ref{tab:2} and \ref{tab:3} it is apparent that 
the differences between the energies resulting from the localised
and original pseudopotentials is negligible in comparison to the 
differences between the energies obtained from the AE and 
pseudopotential calculations. This suggests that the localisation 
procedure has been successful, and that the transferability is 
unlikely to be improved by requiring a more exact reproduction 
of the original pseudo-orbitals.

Our localised pseudopotentials reproduce the AE
excitation energies well, suggesting that our pseudopotentials are
transferable, and that the small relaxation of the norm-conservation
criterion has no significant effect.  Our pseudopotentials also appear
to reproduce the excitation energies rather better than those 
of Ovcharenko \emph{et al.}
In a few cases this appears to be due to the lack of a $d$ channel 
in the pseudopotentials of Ovcharenko \emph{et al.} (e.g. P), but for 
most atoms considered this is not significant (e.g. Al).

\begin{figure}
\includegraphics{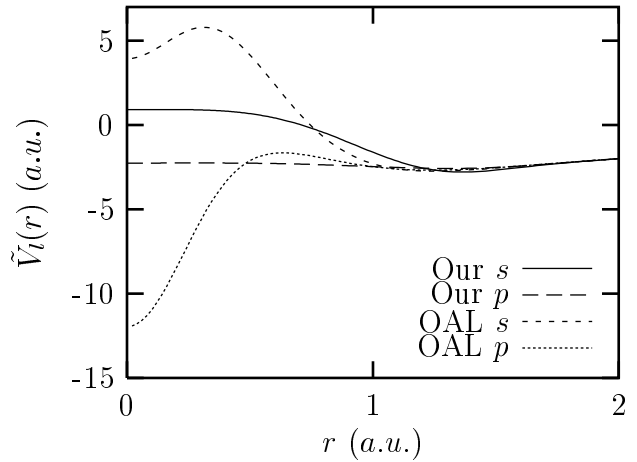}
\caption{\label{fig4} Comparison of the $s$ and $p$ channels of the 
localised Si pseudopotentials generated within this paper (Our) and 
by Ovcharenko
\emph{et al.}\cite{ovcharenko01} (OAL). The OAL pseudopotential does
not have a $d$ channel for us to compare with.}
\end{figure}

In addition to this our pseudopotentials
are softer than those of Ovcharenko \emph{et al.}, see Fig.~\ref{fig4}
for a comparison of the Si pseudopotentials.

\section{Conclusions}
\label{sec:conc}
We have shown that in general a non-local norm-conserving
pseudopotential constructed within HF theory will be non-local over
all space, resulting in a long range interaction between atoms that is
incorrect and not present in the equivalent AE calculation.  In
particular it is apparent that the total-energy cannot be defined for
extended systems.  Although our analysis of extreme non-locality was 
for closed shell atoms within HF theory, it is apparent that 
the same non-locality persists for open shell atoms and 
for Dirac-Fock theory, since the nature of the exchange 
interaction remains the same (although the explicit equations 
become more complex) - it is represented by a non-local functional 
of all states which is different for each state.

This extreme non-locality is of particular relevance to the
application of pseudopotentials within the DMC method.  HF
pseudopotentials have been shown to give good results within DMC
calculations, but computational efficiency requires the region of
non-locality to be as small as possible.  In light of this we have
implemented a localisation procedure which generates a new pseudopotential
that is non-local in a small region surrounding each nucleus.
Localisation together with reproduction of the atomic eigenvalues and 
logarithmic derivatives is achieved at the cost of a small relaxation 
of the norm-conservation condition, due to an alteration of the 
asymptotic form of the pseudo-orbital.

We have calculated localised pseudopotentials for H-Ar, and
tests for these atoms demonstrate the reliability of our approach.
We have also demonstrated their transferability by presenting
calculations of excited states for these atoms.  This suggests that 
the approach given here provides accurate, well localised, smooth, 
HF pseudopotentials suitable for use in \emph{ab initio} methods such 
as DMC, $GW$, and quantum chemistry techniques.

In a future paper we will further generalise our approach to
relativistic HF pseudopotentials (relativistic effective potentials),
parameterise these pseudopotential in terms of a Gaussian basis set for
use in quantum chemistry codes, and present a library of 
pseudopotentials for $Z=1-56$ and $71-80$.

\begin{acknowledgements}
Financial support was provided by the Engineering and Physical
Sciences Research Council (EPSRC), UK.
\end{acknowledgements}

\begin{table*}[t]
\begin{tabular}{llcccc} \hline \hline
Atom\ \ \ \ &
 Configuration\ \ \ \ &
\ \ \ \ $E_{exc}^{AE}$\ \ \ \ &
\ \ \ \ $\tilde{E}_{exc}$ \ \ \ \ &
\ \ \ \ $\tilde{E}_{exc}^{loc}$ \ \ \ \ &
\ \ \ \ $E_{exc}^{OAL}$ \\ \hline
H  & $1s^1        [^2S]$ & 0.0000 & 0.0000 & 0.0000 &   -    \vspace{-0.1cm} \\
   & $2p^1        [^2P]$ & 0.3750 & 0.3750 & 0.3750 &   -    \vspace{-0.1cm} \\
   & $3d^1        [^2D]$ & 0.4444 & 0.4444 & 0.4444 &   -    \vspace{+0.1cm} \\
He & $1s^2        [^1S]$ & 0.0000 & 0.0000 & 0.0000 &   -    \vspace{-0.1cm} \\
   & $1s^12p^1    [^3P]$ & 0.7302 & 0.7308 & 0.7308 &   -    \vspace{-0.1cm} \\
   & $1s^13d^1    [^3D]$ & 0.8061 & 0.8067 & 0.8067 &   -    \vspace{+0.1cm} \\
Li & $2s^1        [^2S]$ & 0.0000 & 0.0000 & 0.0000 &   -    \vspace{-0.1cm} \\
   & $2p^1        [^2P]$ & 0.0677 & 0.0677 & 0.0677 &   -    \vspace{-0.1cm} \\
   & $3d^1        [^2D]$ & 0.1407 & 0.1408 & 0.1408 &   -    \vspace{+0.1cm} \\
Be & $2s^2        [^1S]$ & 0.0000 & 0.0000 & 0.0000 &   -    \vspace{-0.1cm} \\
   & $2s^12p^1    [^3P]$ & 0.0615 & 0.0595 & 0.0595 &   -    \vspace{-0.1cm} \\
   & $2s^13d^1    [^3D]$ & 0.2389 & 0.2390 & 0.2390 &   -    \vspace{+0.1cm} \\
B  & $2s^22p^1    [^2P]$ & 0.0000 & 0.0000 & 0.0000 & 0.0000 \vspace{-0.1cm} \\
   & $2s^12p^2    [^4P]$ & 0.0784 & 0.0764 & 0.0765 & 0.0774 \vspace{-0.1cm} \\
   & $2s^23d^1    [^2D]$ & 0.2353 & 0.2358 & 0.2358 & 0.2356 \vspace{+0.1cm} \\
C  & $2s^22p^2    [^3P]$ & 0.0000 & 0.0000 & 0.0000 & 0.0000 \vspace{-0.1cm} \\
   & $2s^12p^3    [^5S]$ & 0.0894 & 0.0840 & 0.0841 & 0.0836 \vspace{-0.1cm} \\
   & $2s^22p^13d^1[^3F]$ & 0.3402 & 0.3408 & 0.3407 & 0.3407 \vspace{+0.1cm} \\
N  & $2s^22p^3    [^4S]$ & 0.0000 & 0.0000 & 0.0000 & 0.0000 \vspace{-0.1cm} \\
   & $2s^12p^4    [^4P]$ & 0.4126 & 0.4097 & 0.4095 & 0.4093 \vspace{-0.1cm} \\
   & $2s^22p^23d^1[^4F]$ & 0.4565 & 0.4571 & 0.4570 & 0.4573 \vspace{+0.1cm} \\
O  & $2s^22p^4    [^3P]$ & 0.0000 & 0.0000 & 0.0000 & 0.0000 \vspace{-0.1cm} \\
   & $2s^22p^33d^1[^5D]$ & 0.3809 & 0.3808 & 0.3808 & 0.3809 \vspace{-0.1cm} \\
   & $2s^12p^5    [^3P]$ & 0.6255 & 0.6257 & 0.6256 & 0.6254 \vspace{+0.1cm} \\
F  & $2s^22p^5    [^2P]$ & 0.0000 & 0.0000 & 0.0000 & 0.0000 \vspace{-0.1cm} \\
   & $2s^22p^43d^1[^4F]$ & 0.5220 & 0.5223 & 0.5222 & 0.5211 \vspace{-0.1cm} \\
   & $2s^12p^6    [^2S]$ & 0.8781 & 0.8832 & 0.8830 & 0.8856 \vspace{+0.1cm} \\
Ne & $2s^22p^6    [^1S]$ & 0.0000 & 0.0000 & 0.0000 & 0.0000 \vspace{-0.1cm} \\
   & $2s^22p^53d^1[^3F]$ & 0.6734 & 0.6740 & 0.6738 & 0.6754 \vspace{-0.1cm} \\
   & $2s^12p^63d^1[^3D]$ & 1.7565 & 1.7613 & 1.7610 & 1.7636 \vspace{+0.1cm} \\
Na & $3s^1        [^2S]$ & 0.0000 & 0.0000 & 0.0000 &   -    \vspace{-0.1cm} \\
   & $3p^1        [^2P]$ & 0.0725 & 0.0727 & 0.0727 &   -    \vspace{-0.1cm} \\
   & $3d^1        [^2D]$ & 0.1263 & 0.1264 & 0.1264 &   -    \vspace{+0.1cm} \\
Mg & $3s^2        [^1S]$ & 0.0000 & 0.0000 & 0.0000 &   -    \vspace{-0.1cm} \\
   & $3s^13p^1    [^3P]$ & 0.0679 & 0.0672 & 0.0671 &   -    \vspace{-0.1cm} \\
   & $3s^13d^1    [^3D]$ & 0.1843 & 0.1845 & 0.1845 &   -    \vspace{+0.1cm} \\
Al & $3s^23p^1    [^2P]$ & 0.0000 & 0.0000 & 0.0000 & 0.0000 \vspace{-0.1cm} \\
   & $3s^13p^2    [^4P]$ & 0.0858 & 0.0850 & 0.0851 & 0.0758 \vspace{-0.1cm} \\
   & $3s^23d^1    [^2D]$ & 0.1441 & 0.1443 & 0.1443 & 0.1478 \vspace{+0.1cm} \\
Si & $3s^23p^2    [^3P]$ & 0.0000 & 0.0000 & 0.0000 & 0.0000 \vspace{-0.1cm} \\
   & $3s^13p^3    [^5S]$ & 0.0913 & 0.0890 & 0.0891 & 0.0884 \vspace{-0.1cm} \\
   & $3s^23p^13d^1[^3F]$ & 0.2146 & 0.2147 & 0.2147 & 0.2185 \vspace{+0.1cm} \\
P  & $3s^23p^3    [^4S]$ & 0.0000 & 0.0000 & 0.0000 & 0.0000 \vspace{-0.1cm} \\
   & $3s^23p^23d^1[^4F]$ & 0.3006 & 0.3010 & 0.3009 & 0.3059 \vspace{-0.1cm} \\
   & $3s^13p^4    [^4P]$ & 0.3023 & 0.3030 & 0.3030 & 0.3029 \vspace{+0.1cm} \\
S  & $3s^23p^4    [^3P]$ & 0.0000 & 0.0000 & 0.0000 & 0.0000 \vspace{-0.1cm} \\
   & $3s^23p^33d^1[^5D]$ & 0.2671 & 0.2658 & 0.2657 & 0.2707 \vspace{-0.1cm} \\
   & $3s^13p^5    [^3P]$ & 0.4260 & 0.4269 & 0.4269 & 0.4257 \vspace{+0.1cm} \\
Cl & $3s^23p^5    [^2P]$ & 0.0000 & 0.0000 & 0.0000 & 0.0000 \vspace{-0.1cm} \\
   & $3s^23p^43d^1[^4F]$ & 0.3733 & 0.3735 & 0.3734 & 0.3754 \vspace{-0.1cm} \\
   & $3s^13p^6    [^2S]$ & 0.5653 & 0.5671 & 0.5671 & 0.5675 \vspace{+0.1cm} \\
Ar & $3s^23p^6    [^1S]$ & 0.0000 & 0.0000 & 0.0000 & 0.0000 \vspace{-0.1cm} \\
   & $3s^23p^53d^1[^3F]$ & 0.4824 & 0.4838 & 0.4837 & 0.4855 \vspace{-0.1cm} \\
   & $3s^13p^63d^1[^3D]$ & 1.1597 & 1.1623 & 1.1621 & 1.1643 \vspace{+0.1cm} \\ \hline
\multicolumn{3}{l}{Average error} & $+1\times10^{-4}$ 
                                  & $+7\times10^{-5}$ 
                                  & $+9\times10^{-4}$  \\ 
\multicolumn{3}{l}{Maximum error} & $-5\times10^{-3}$ 
                                  & $-5\times10^{-3}$ 
                                  & $-1\times10^{-2}$  \\ \hline \hline
\end{tabular}
\caption{Comparison of excitation energies obtained from AE 
calculations ($E_{exc}^{AE}$), from calculations using the 
extremely non-local and localised pseudopotentials generated 
within this paper ($\tilde{E}_{exc}$ and $\tilde{E}_{exc}^{loc}$ 
respectively), and from calculations using the pseudopotentials 
generated by Ovcharenko \emph{et al.}\cite{ovcharenko01} 
($E_{exc}^{OAL}$) (a.u.). }
\label{tab:2}
\end{table*}

\begin{table}[t]
\begin{tabular}{lcccc} \hline \hline
Atom\ \ \ \ &
\ \ \ \ $E_{ion}^{AE}$\ \ \ \ &
\ \ \ \ $\tilde{E}_{ion}$ \ \ \ \ &
\ \ \ \ $\tilde{E}_{ion}^{loc}$ \ \ \ \ &
\ \ \ \ $E_{ion}^{OAL}$ \\ \hline
H  &  0.50000  &  0.50000  &  0.50000  &     -    \\
He &  0.86168  &  0.86230  &  0.86230  &     -    \\
Li &  0.19631  &  0.19632  &  0.19632  &     -    \\
Be &  0.29563  &  0.29578  &  0.29578  &     -    \\
B  &  0.29149  &  0.29199  &  0.29196  &  0.29173 \\
C  &  0.39640  &  0.39702  &  0.39693  &  0.39696 \\
N  &  0.51293  &  0.51357  &  0.51352  &  0.51378 \\
O  &  0.43679  &  0.43672  &  0.43680  &  0.43679 \\
F  &  0.57763  &  0.57790  &  0.57794  &  0.57672 \\
Ne &  0.72928  &  0.72982  &  0.72982  &  0.73124 \\
Na &  0.18195  &  0.18210  &  0.18210  &     -    \\
Mg &  0.24283  &  0.24299  &  0.24303  &     -    \\
Al &  0.20204  &  0.20227  &  0.20227  &  0.20531 \\
Si &  0.28123  &  0.28175  &  0.28172  &  0.28164 \\
P  &  0.36901  &  0.36984  &  0.36977  &  0.36957 \\
S  &  0.33171  &  0.33061  &  0.33060  &  0.33130 \\
Cl &  0.43348  &  0.43372  &  0.43366  &  0.43339 \\
Ar &  0.54298  &  0.54444  &  0.54433  &  0.54395 \\ \hline
\multicolumn{2}{l}{Average error} & $ 3\times10^{-4}$ 
                                  & $ 3\times10^{-4}$ 
                                  & $ 6\times10^{-4}$  \\ 
\multicolumn{2}{l}{Maximum error} & $ 1\times10^{-3}$ 
                                  & $ 1\times10^{-3}$ 
                                  & $-3\times10^{-3}$  \\ \hline \hline
\end{tabular}
\caption{Comparison of ionisation potentials obtained from AE
calculations ($E_{ion}^{AE}$), from calculations using the
extremely non-local and localised pseudopotentials generated
within this paper ($\tilde{E}_{ion}$ and $\tilde{E}_{ion}^{loc}$
respectively), and from calculations using the pseudopotentials
generated by Ovcharenko \emph{et al.}\cite{ovcharenko01}
($E_{ion}^{OAL}$) (a.u.).  }
\label{tab:3}
\end{table}



\begin{thebibliography}{}
\bibitem{nl-ext} DFT is only strictly valid for local external
potentials, hence for a non-local pseudopotential there is formally no
one-to-one correspondence between the charge density and the (now
non-local) effective potential.  However, it is common practice to
choose a particular correspondence in order to implement KS-DFT with 
non-local pseudopotentials. In standard KS-DFT we define the
effective potential as the sum of the non-local pseudopotential and
the same explicit functional of density used in the local external
potential case.  For the OPM method we define the effective potential
as the sum of the non-local pseudopotential and the same implicit
functional of density used in the local external potential case.
\end{thebibliography}
\end{document}